\documentclass{emulateapj}

\newcommand{\ajp}{Aust.~J.~Phys.}              
\newcommand{\cpl}{Chem.~Phys.~Lett.}           
\newcommand{\cjp}{Can.~J.~Phys.}               
\newcommand{\jpcrd}{J.~Phys.~Chem.~Ref.~Data}  
\newcommand{\zfpd}{Zeit.~Phys.~D: Atoms, Molecules and Clusters}  %

\shorttitle{The chemistry of vibrationally excited H$_2$}
\shortauthors{Ag\'undez et al.}

\begin{document}

\title{The chemistry of vibrationally excited H$_2$ in the interstellar medium}

\author{M. Ag\'undez\altaffilmark{1}, J. R. Goicoechea\altaffilmark{2}, J. Cernicharo\altaffilmark{2}, A. Faure\altaffilmark{3}, and E. Roueff\altaffilmark{1}}

\altaffiltext{1}{LUTH, Observatoire de Paris-Meudon, 5 Place Jules
Janssen, 92190 Meudon, France; marcelino.agundez@obspm.fr,
evelyne.roueff@obspm.fr}

\altaffiltext{2}{Departamento de Astrof\'isica, Centro de
Astrobiolog\'ia, CSIC-INTA, Ctra. de Torrej\'on a Ajalvir km 4,
Torrej\'on de Ardoz, 28850 Madrid, Spain;
jr.goicoechea@cab.inta-csic.es, jcernicharo@cab.inta-csic.es}

\altaffiltext{3}{Laboratoire d'Astrophysique de Grenoble (LAOG),
Universit\'e Joseph Fourier, UMR 5571 CNRS, BP 53, 38041 Grenoble
cedex 09, France; afaure@obs.ujf-grenoble.fr}

\begin{abstract}
The internal energy available in vibrationally excited H$_2$
molecules can be used to overcome or diminish the activation
barrier of various chemical reactions of interest for molecular
astrophysics. In this article we investigate in detail the impact
on the chemical composition of interstellar clouds of the
reactions of vibrationally excited H$_2$ with C$^+$, He$^+$, O,
OH, and CN, based on the available chemical kinetics data. It is
found that the reaction of H$_2$ ($v>0$) and C$^+$ has a profound
impact on the abundances of some molecules, especially CH$^+$,
which is a direct product and is readily formed in astronomical
regions with fractional abundances of vibrationally excited H$_2$,
relative to ground state H$_2$, in excess of $\sim$ 10$^{-6}$,
independently of whether the gas is hot or not. The effects of
these reactions on the chemical composition of the diffuse clouds
$\zeta$ Oph and HD 34078, the dense PDR Orion Bar, the planetary
nebula NGC 7027, and the circumstellar disk around the B9 star HD
176386 are investigated through PDR models. We find that formation
of CH$^+$ is especially favored in dense and highly FUV
illuminated regions such as the Orion Bar and the planetary nebula
NGC 7027, where column densities in excess of 10$^{13}$ cm$^{-2}$
are predicted. In diffuse clouds, however, this mechanism is found
to be not efficient enough to form CH$^+$ with a column density
close to the values derived from astronomical observations.
\end{abstract}



\section{Introduction}

As concerns the chemical modelling of interstellar clouds,
state-to-state chemistry, i.e. that which considers the quantum
states of reactants and/or products in a given chemical reaction,
has been traditionally ignored, mainly because the required
chemical kinetics data are generally not available and because it
is expected that it would not affect in a drastic way the
predicted chemical abundances. Some attention however has been
given to the different reactivity of ortho and para states of
molecular hydrogen and its effect on various aspects of
interstellar chemistry such as the formation of ammonia or the
ortho-to-para ratio of c-C$_3$H$_2$ in cold dense clouds
\citep{leb91,par06}. The enhancement in the reactivity of
molecular hydrogen when it is in a vibrationally excited state has
also been considered in the past, mainly in the context of trying
to explain the long standing problem of CH$^+$ formation in
diffuse clouds \citep{ste72,fre82,gar03}, but also in the study of
the chemistry of dense photon dominated regions (PDRs)
\citep{tie85,ste95} and shocked regions \citep{wag87,hol89}. These
studies assumed different reactivity enhancements for reactions of
vibrationally excited H$_2$ with species such as C$^+$, O, or C,
and found that whereas in the case of diffuse clouds a too large
fraction of H$_2$ $v>0$ is required to explain the formation of
CH$^+$ with the observed abundance, in the case of dense PDRs
there are important effects for the chemistry.

In this paper we look again at the chemistry of vibrationally
excited H$_2$ in the light of some experimental and theoretical
studies carried out in the last years, which have been essentially
ignored by the astrochemical community, and investigate the
implications for various astronomical regions where a relatively
large fraction of vibrationally excited H$_2$ has been observed or
is expected to be present.

\begin{deluxetable*}{rllcrr}
\tabletypesize{\scriptsize} \tablecaption{Thermal and
state-specific rate constants for chemical reactions of H$_2$
relevant for astrophysics\label{table-h2-reac}} \tablewidth{0pc}
\startdata \hline
\multicolumn{1}{c}{No} & \multicolumn{1}{c}{Reaction} & \multicolumn{1}{c}{$k$ (cm$^3$ s$^{-1}$)} & \multicolumn{1}{c}{$\Delta T$ (K)\tablenotemark{a}} & \multicolumn{1}{c}{$\Delta H_r^0(0K)$ (K)} & \multicolumn{1}{c}{Reference} \\
\hline
1  & H$_2$ + C$^+$ $\rightarrow$ CH$^+$ + H             & 7.4$\times$10$^{-10}$ $\exp(-4537/T)$                 & 400-1300 & $+$4280   & (1) \\
2  & H$_2$($j=0,7$) + C$^+$ $\rightarrow$ CH$^+$ + H     & 1.58$\times$10$^{-10}$ $\exp(-[4827-E_j/k]/T)$\tablenotemark{b}      & 200-1000 & $(+4280,-310)$   & (2) \\
3  & H$_2$($v=1$) + C$^+$ $\rightarrow$ CH$^+$ + H      & 1.6$\times$10$^{-9}$                                    & 800-1300 & $-$1710   & (1) \\
\hline
4  & H$_2$ + He$^+$ $\rightarrow$ He + H + H$^+$    & 3.7$\times$10$^{-14}$ $\exp(-35/T)$                   & 10-300   & $-$75560  & (3) \\
5  & H$_2$($v>1$) + He$^+$ $\rightarrow$ He + H + H$^+$ & 0.18-1.8$\times$10$^{-9}$                               & 300      & $-$87190  & (4) \\
\hline 11 & H$_2$ + O $\rightarrow$ OH + H                 & 3.52$\times$10$^{-13}$ $(T/300)^{2.60} \exp(-3241/T)$ & 297-3532 & $+$920    & (5) \\
6  & H$_2$($v=1$) + O $\rightarrow$ OH + H              & 1.68$\times$10$^{-16}$ $(T/300)^{9.34} \exp(943/T)$   & 100-500\tablenotemark{c} & $-$5070   & (6) \\
7  & H$_2$($v=2$) + O $\rightarrow$ OH + H              & 1.52$\times$10$^{-13}$ $(T/300)^{5.13} \exp(209/T)$   & 100-500\tablenotemark{c} & $-$10720  & (6) \\
8  & H$_2$($v=3$) + O $\rightarrow$ OH + H              & 2.07$\times$10$^{-11}$ $(T/300)^{0.98} \exp(-412/T)$  & 100-4000 & $-$16040  & (6) \\
\hline
9  & H$_2$ + OH $\rightarrow$ H$_2$O + H            & 2.22$\times$10$^{-12}$ $(T/300)^{1.43} \exp(-1751/T)$ & 200-3000 & $-$7370   & (5) \\
10 & H$_2$($v=1$) + OH $\rightarrow$ H$_2$O + H         & 1.52$\times$10$^{-11}$ $(T/300)^{1.33} \exp(-902/T)$  & 250-2000 & $-$13360  & (7) \\
\hline
11 & H$_2$ + CN $\rightarrow$ HCN + H               & 1.17$\times$10$^{-12}$ $(T/300)^{2.31} \exp(-1188/T)$ & 200-3500 & $-$10250  & (5) \\
12 & H$_2$($v=1$) + CN $\rightarrow$ HCN + H            &
9.65$\times$10$^{-12}$ $(T/300)^{1.04} \exp(-1397/T)$ & 200-1000 &
$-$16240  & (8)
\enddata
\tablecomments{When no specific state of H$_2$ is indicated the
expression for $k$ corresponds to the thermal rate constant. The
reaction enthalpies $\Delta H_r^0(0K)$ have been computed from the
formation enthalpies at 0~K of each species, taken from NIST-JANAF
Thermochemical Tables \citep{cha98}, while for the H$_2$
state-specific reactions the level energies of H$_2$ have been
taken from \citet{dab84}.} \tablenotetext{a}{$\Delta T$ is the
temperature range over which the rate constant has been
studied.}\tablenotetext{b}{$E_j$ is the energy of each rotational
level of H$_2$ ($v=0,j$) relative to the ground state $v=0,j=0$
($E_1$=170.5 K, $E_2$=509.9 K, $E_3$=1015.2 K, $E_4$=1681.7 K,
$E_5$=2503.9 K, $E_6$=3474.4 K, and $E_7$=4586.4 K;
\citealt{dab84}).}\tablenotetext{c}{In the temperature range
500-4000~K the rate constants for the reactions between O and
H$_2$ in the $v=1$ and $v=2$ states can be better fit by the
expressions 7.79$\times$10$^{-12}$ ($T$/300)$^{1.20}$
$\exp$(-2444/$T$) and 2.39$\times$10$^{-11}$ ($T$/300)$^{0.87}$
$\exp$(-1325/$T$) respectively.} \tablerefs{(1) \citet{hie97}; (2)
based on \citet{ger87}; (3) UMIST Database for Astrochemistry
\citep{woo07}; (4) \citet{jon86}; (5) NIST Chemical Kinetics
Database (\texttt{http://kinetics.nist.gov/kinetics/}); (6)
\citet{sul05}; (7) \citet{zel81}, \citet{tru95}; (8)
\citet{zhu98}.}
\end{deluxetable*}

\section{The chemistry of vibrationally excited H$_2$}

\subsection{Basics} \label{subsec-h2chem-basics}

\begin{figure}
\includegraphics[angle=0,scale=.61]{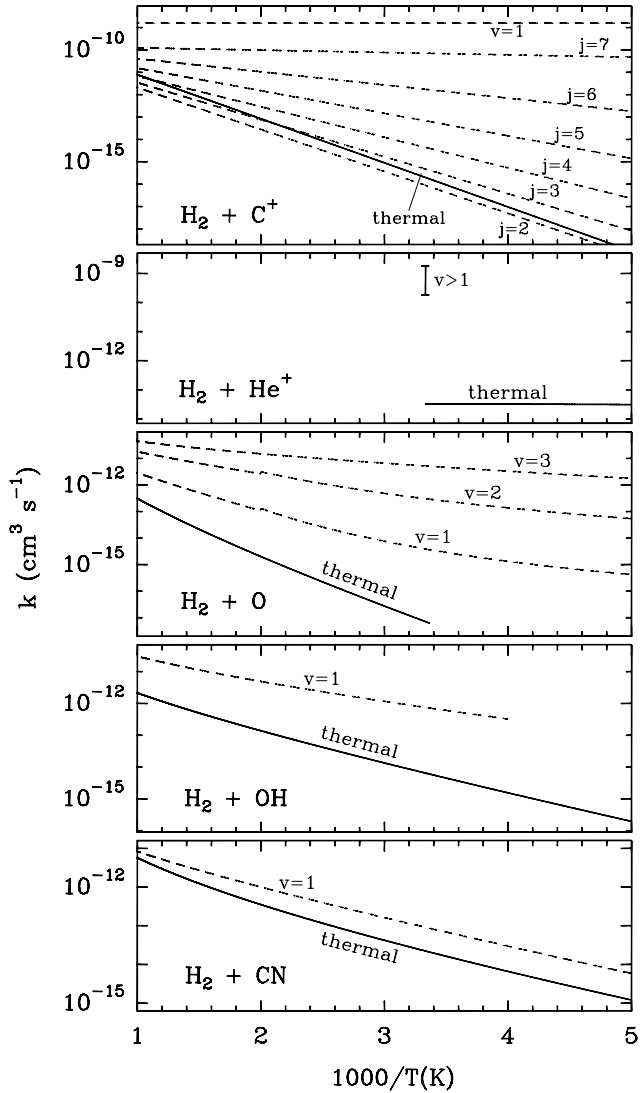}
\caption{Rate constants as a function of the reciprocal of
temperature for thermal and state-specific reactions of H$_2$ with
C$^+$, He$^+$, O, OH, and CN.} \label{fig-krates}
\end{figure}

The main feature that characterizes the chemistry of H$_2$ in
excited vibrational states -as compared to that of H$_2$ in the
ground vibrational state- is that the internal energy of the
excited hydrogen molecule can be used to overcome or diminish
activation barriers which are present when H$_2$ is in its ground
vibrational state\footnote{In regard to the chemistry of
vibrationally excited H$_2$ it is worth to comment on the
difference between a state-specific and a thermal rate constant.
The thermal rate constant, usually provided by experimental and
theoretical chemists and used in interstellar chemical models, is
the thermal average of the individual state-specific values, i.e.
the sum over all H$_2$ states of the state-specific rate constant
multiplied by the thermal fractional population of that H$_2$
state. It turns out that since the vibrationally excited levels of
H$_2$ are very high in energy and thus poorly populated in thermal
conditions, except at high temperatures ($>$ 1000~K), we expect
the thermal rate constant to be essentially equal to the
state-specific H$_2$ $v=0$ rate constant, so that hereafter we use
both terms interchangeably.}. Several experimental and theoretical
studies have demonstrated that such behavior occurs for some
reactions of H$_2$ which are of relevance for astrophysics, the
kinetic data of which are detailed in Table~\ref{table-h2-reac}
and Fig.~\ref{fig-krates}.

The first reaction that draw our attention is
\begin{equation}
\rm H_2 + C^+ \rightarrow CH^+ + H \label{reac-h2+cplus}
\end{equation}
which is well known to possess an endothermicity of 0.37 eV
($\sim$ 4300~K). It has long been thought that such endothermicity
could be overcome by vibrationally excited H$_2$ (the $v=1$ levels
have energies $>$ 0.5 eV above the ground state), hypothesis that
gained experimental support with the study of \citet{jon86} and
the more recent measurements of \citet{hie97}. These latter
authors were able to measure the rate constant of H$_2$($v=1$) +
C$^+$ over the 800-1300~K temperature range and found it to be in
the range 1-2$\times$10$^{-9}$ cm$^3$ s$^{-1}$, i.e. essentially
equal to the Langevin collision rate (1.6$\times$10$^{-9}$ cm$^3$
s$^{-1}$), and to show a slight negative temperature dependence.
This is the typical behavior of exothermic ion-non polar neutral
reactions which have a temperature independent rate constant close
to the collision limit, and indicates that it is quite safe to
adopt also the collision rate at low temperatures for the reaction
of H$_2$($v=1$) and C$^+$. In connection with this we note that
rate coefficients for reaction (\ref{reac-h2+cplus}) have been
computed by \citet{ger87} for selected initial rotational levels
($j=0,7$) of H$_2$ in the ground vibrational state. This work was
combined with guided-beam measurements and a good agreement
between the experimental results and the statistical calculations
was observed. \citet{ger87} found that the state-specific rate
constants may be well approximated by Arrhenius expressions (see
Table~\ref{table-h2-reac} and Fig.~\ref{fig-krates}) in which as
$j$ increases the exponential factor decreases by an amount equal
to the energy of the $j$ level. All this indicates that the energy
of excited H$_2$ in reaction (\ref{reac-h2+cplus}) is effectively
used to diminish or overcome the endothermicity of the reaction,
and that for H$_2$ states with $v>0$ the reaction proceeds at the
collision rate.

The second reaction in which we focus is
\begin{equation}
\rm H_2 + He^+ \rightarrow He + H + H^+ \label{reac-h2+heplus}
\end{equation}
which is strongly exothermic but shows an extremely small rate
constant, some 10$^{-14}$ cm$^3$ s$^{-1}$ below 300~K
\citep{sch89}, which indicates the presence of an important
activation barrier. \citet{jon86} have studied the kinetics of
this reaction experimentally at room temperature and found that
when H$_2$ is in a vibrational state $v>1$ the reaction proceeds
at about the collision limit, 1.8$\times$10$^{-9}$ cm$^3$
s$^{-1}$, value that we adopt hereafter.

Another interesting reaction is
\begin{equation}
\rm H_2 + O \rightarrow OH + H \label{reac-h2+o}
\end{equation}
which is known to be endothermic by 0.08 eV ($\sim$ 900~K) and to
possess an even higher activation barrier of about 0.4 eV ($\sim$
4800 K). Recently \citet{sul05} have calculated the rate constants
for reaction (\ref{reac-h2+o}) with H$_2$ in various vibrational
states ($v=0-3$) and have found that although the rate constant is
noticeably enhanced as the vibrational state of H$_2$ is
increased, the reaction does not proceed at the collision rate
limit for H$_2$ ($v>0$) but an activation barrier persists even
when H$_2$ is in the $v=3$ state (see Fig.~\ref{fig-krates}).

The exothermic reaction
\begin{equation}
\rm H_2 + OH \rightarrow H_2O + H \label{reac-h2+oh}
\end{equation}
possesses an activation barrier which can be partially overcome,
when H$_2$ is in first excited vibrational state
(\citealt{zel81,tru95}, see Fig.~\ref{fig-krates}). Something
similar occurs for the exothermic reaction
\begin{equation}
\rm H_2 + CN \rightarrow HCN + H \label{reac-h2+cn}
\end{equation}
the rate constant of which experiences a moderate enhancement when
H$_2$ is in the vibrational state $v=1$ (\citealt{zhu98}, see
Fig.~\ref{fig-krates}).\\

The studies carried out on the reactions commented above leave us
the following general picture. The reactions of H$_2$ which are
endothermic or possess activation barriers occur faster as the
internal excitation of H$_2$ increases. In some cases the internal
energy of excited H$_2$ is effectively used to overcome the
reaction barrier and the rate constant approaches the collision
limit, whereas in other cases an activation barrier persists even
if the internal energy of excited H$_2$ makes the reaction to be
exothermic. The rate constant enhancement is therefore quite
specific of each reaction and difficult to predict\footnote{We
also note that for fast exothermic reactions (without barriers),
the internal excitation of a reactant can actually inhibit the
reactivity, as observed experimentally for ion-molecule (e.g.
\citealt{ger89}) and neutral-neutral \citep{olk07} reactions.}. In
particular, it is not simply given by the subtraction of the
internal energy of excited H$_2$ from the endothermicity or
activation barrier of the reaction with ground state H$_2$, as has
been assumed previously (e.g. \citealt{ste95}). The rate constant
enhancement when H$_2$ goes from the ground to an excited
vibrational state gets larger at low temperatures and can reach
many orders of magnitude. For instance, at 200~K it is somewhat
less than a factor of ten for the reaction with CN, whereas for
H$_2$ + C$^+$ it is as high as ten orders of magnitude.

The state-specific rate constants for reactions of H$_2$ are
obviously useless to study a gas in which the H$_2$ states are
thermally populated, in which case the thermal rate constants are
to be used, but become very useful for the study of regions where
H$_2$ levels have non thermal populations, which is found to occur
quite often in the interstellar medium. In such regions some
state-specific reactions of vibrationally excited H$_2$ may become
important to determine the chemical composition of the gas.

Let us consider a region with a gas kinetic temperature $T$ where
H$_2$ levels are not thermally populated, so that the fractional
abundance of vibrationally excited H$_2$ is $f^*$ = $n$(H$_2$
$v>0$)/$n$(H$_2$ $v=0$), and the generic reactions
\begin{equation}
{\rm H_2}(v=0) + {\rm X} \stackrel{k^0(T)}{\longrightarrow} {\rm
products} \label{reac-h2v0+x}
\end{equation}
\begin{equation}
{\rm H_2}(v>0) + {\rm X} \stackrel{k^*(T)}{\longrightarrow} {\rm
products} \label{reac-h2vexc+x}
\end{equation}
so that $\Delta k(T)$ = $k^*(T)/k^0(T)$ is the rate constant
enhancement at a temperature $T$. For reaction
(\ref{reac-h2vexc+x}) to become important it has to be competitive
with respect to reaction (\ref{reac-h2v0+x}), i.e. $f^* \Delta
k(T) > 1$, and also with respect to other reactions which consume
the reactant X. According to this, reactions such as H$_2$ + CN,
which has a moderate rate enhancement ($\Delta k \lesssim 10$),
need a very high abundance of vibrationally excited H$_2$ ($f^* >
0.1$) to have some impact on the chemistry. On the other hand,
reactions with a very high rate constant enhancement, such as that
of H$_2$ and C$^+$ for which $\Delta k \sim 10^{10}$ at 200~K,
just need a moderate abundance of vibrationally excited H$_2$
($f^* > 10^{-10}$). This is however a condition necessary but not
sufficient to have some impact on the chemistry as the reaction
rate of, in this case, H$_2$($v>0$) + C$^+$ needs also to
be large enough compared to other reactions which consume C$^+$.\\

\begin{figure*}
\includegraphics[angle=-90,scale=.71]{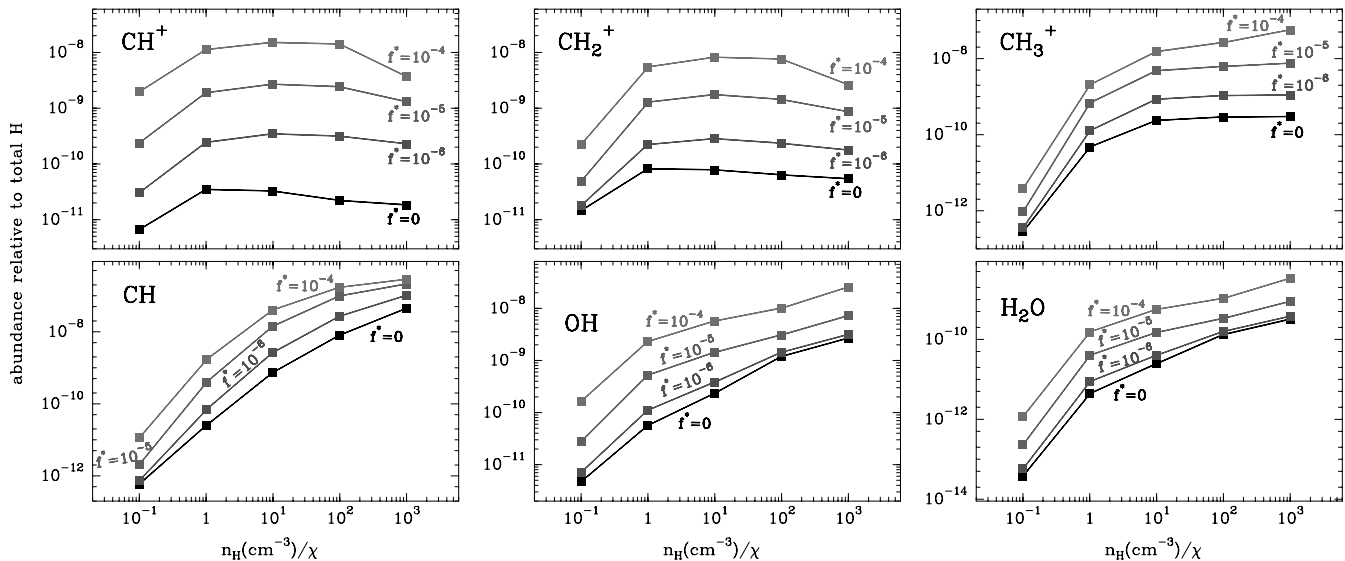}
\caption{Steady state abundances relative to total H nuclei for
CH$^+$, CH$_2^+$, CH$_3^+$, CH, OH, and H$_2$O as a function of
$n_{\rm H}/\chi$ for various fractional abundances $f^*$ of H$_2$
($v=1$). All the models have been run with $n_{\rm H}$ = 10$^3$
cm$^{-3}$ (so that the FUV field strength $\chi$ has been varied),
$A_V$ = 0.5, and $T$ = 100~K.} \label{fig-abun}
\end{figure*}

\subsection{Implications for the interstellar medium}

We may state that the importance of the chemistry of vibrationally
excited H$_2$ is limited to reactions with a large rate constant
enhancement and to astronomical regions with a high abundance of
vibrationally excited H$_2$. In order to get a quantitative
estimate of the impact of these reactions on the chemistry of the
interstellar medium we have run single point time dependent
chemical models covering a wide range of the space of parameters:
fraction of vibrationally excited H$_2$ ($f^*$), density of
hydrogen nuclei ($n_{\rm H}$), far ultraviolet (FUV) radiation
field strength ($\chi$, relative to the Draine interstellar
radiation field; \citealt{dra78}), and gas kinetic temperature
($T$).

We consider 127 different species composed of the elements H, He,
C, N, O, and S, whose abundances are assumed to be solar
\citep{asp09} with a 25 \% of depletion for elements heavier than
He, and assume that initially all the elements are in atomic form,
either neutral or ionized. The chemical network comprises 2307 gas
phase reactions whose rate constants have been mostly taken from
\citet{agu06} and from the UMIST database for
astrochemistry\footnote{See \texttt{http://www.udfa.net/}}
\citep{woo07}, some of them revised according to the most recent
literature on chemical kinetics. The photodissociation rates of
H$_2$ and CO have been taken from \citet{lee96}. The adopted
visual extinction is 0.5 mag and the cosmic-ray ionization rate
per H$_2$ molecule is 5$\times$10$^{-17}$ s$^{-1}$. We do not
consider grain surface reactions except for the formation of
H$_2$, which is assumed to occur with a rate constant of
3$\times$10$^{-17}$ cm$^3$ s$^{-1}$, the canonical value in
typical diffuse clouds \citep{jur75}.

This quite standard chemical network is completed by adding the
reactions of vibrationally excited H$_2$ detailed in
Table~\ref{table-h2-reac}. Thus, H$_2$ ($v=1$) is included as a
new chemical species with a fractional abundance $f^*$, relative
to ground state H$_2$, which is kept fixed at four different
values (0, 10$^{-6}$, 10$^{-5}$, and 10$^{-4}$), independently of
the mechanism responsible of the excitation of H$_2$. The amount
of H$_2$ ($v>1$) is important for reactions (\ref{reac-h2+heplus})
and (\ref{reac-h2+o}) so that we also consider H$_2$ in the $v=2$
and $v=3$ states with populations which are 1/3 and 1/9
respectively of that in the $v=1$ state, values typically attained
in interstellar clouds with an important fraction of vibrationally
excited H$_2$ (e.g. \citealt{van86,mey01}). In these exploratory
models we do not distinguish between the different reactivity of
H$_2$ in the various rotational levels of the ground vibrational
state.

The abundance of various species are found to be sensitive to the
fraction of vibrationally excited H$_2$. In Fig.~\ref{fig-abun} we
plot the steady state abundance of some of these species for
various fractional abundances of H$_2$ ($v=1$) as a function of
the $n_{\rm H}/\chi$ ratio. The steady state abundances are found
to depend on this latter ratio rather than on either $n_{\rm H}$
or $\chi$. For the results shown in Fig.~\ref{fig-abun} a density
of hydrogen nuclei of 10$^3$ cm$^{-3}$ has been adopted, so that
the FUV field strength has been varied. Reactions of vibrationally
excited H$_2$ have a larger impact on the chemistry at low
temperatures, since as discussed previously the rate constant
enhancement $\Delta k$ of the various reactions is larger. The
results shown in Fig.~\ref{fig-abun} correspond to a relatively
low temperature of 100~K.

The molecule whose abundance is more affected by the reactions of
vibrationally excited H$_2$ is the methylidyne cation (CH$^+$),
which for fractional abundances of H$_2$($v=1$) above
$10^{-7}-10^{-6}$ is readily formed by the reaction between
H$_2$($v=1$) and C$^+$. The steady state abundance of CH$^+$
approximately scales with the fraction of H$_2$($v=1$) and reaches
its maximum abundance for $n_{\rm H}/\chi$ ratios in the range
1$-$100. By running models at different kinetic temperatures we
find that above 400$-$500~K the reaction between H$_2$($v=0$) and
C$^+$ starts to proceed and the abundance of CH$^+$ depends to a
much lesser extent on the fraction of vibrationally excited H$_2$.

The abundances of CH$_2^+$ and CH$_3^+$ are strongly linked to
that of CH$^+$ by the chain of exothermic reactions
\begin{equation}
\rm CH^+ \stackrel{H_2}{\rightarrow} CH_2^+
\stackrel{H_2}{\rightarrow} CH_3^+ \stackrel{H_2}{\nrightarrow}
\end{equation}
which stops at CH$_3^+$, as this species does not react fast with
H$_2$. Therefore, in those conditions in which CH$^+$ reaches a
relatively high abundance, CH$_2^+$ and CH$_3^+$ do so (see
Fig.~\ref{fig-abun}). Some other species such as CH, OH, and
H$_2$O also experience a moderate abundance enhancement when
CH$^+$ is formed. The radical CH is formed by the dissociative
recombination with electrons of the cations CH$_2^+$ and CH$_3^+$,
while OH and H$_2$O are produced by a sequence of reactions that
starts with the photodissociation of CH$^+$ and production of
H$^+$ followed by
\begin{equation}
\rm O \stackrel{H^+}{\rightarrow} O^+ \stackrel{H_2}{\rightarrow}
OH^+ \stackrel{H_2}{\rightarrow} H_2O^+
\stackrel{H_2}{\rightarrow} H_3O^+ \stackrel{e^-}{\rightarrow} OH,
H_2O
\end{equation}

The reaction of vibrationally excited H$_2$ and atomic oxygen has
little effect on the abundance of OH for most of the conditions
explored. It does contribute to OH formation just for fractional
abundances of H$_2$($v=1$) in excess of 10$^{-4}$ and for kinetic
temperatures around 300~K. For much lower temperatures the
activation barriers of the reactions H$_2$($v>0$) + O prevent to
reach a high enough OH formation rate whereas for much larger
temperatures the rate constant enhancements $\Delta k$ are low and
OH is mostly formed by the reaction of H$_2$($v=0$) + O. As
concerns the reactions of vibrationally excited H$_2$ with He$^+$,
OH, and CN, they have almost no impact on the chemistry as the
rate constant enhancements are just moderate and require of a very
large fraction of vibrationally excited H$_2$ to compete with the
corresponding reactions of H$_2$($v=0$).

In summary, from the five reactions of vibrationally excited H$_2$
considered here, only that with C$^+$ has a large impact on the
abundances of some species, especially CH$^+$, in those
astronomical regions which are not too warm (with kinetic
temperatures below a few hundreds of degrees Kelvin) and which
have a moderately high fraction of vibrationally excited H$_2$ (in
excess of $10^{-7}-10^{-6}$). Reactions of H$_2$($v>0$) with
He$^+$, O, OH, and CN are less likely to affect the chemistry of
interstellar clouds unless large amounts of vibrationally excited
H$_2$ are present.

\section{Effects on selected astronomical sources}

\begin{figure}
\includegraphics[angle=-90,scale=.46]{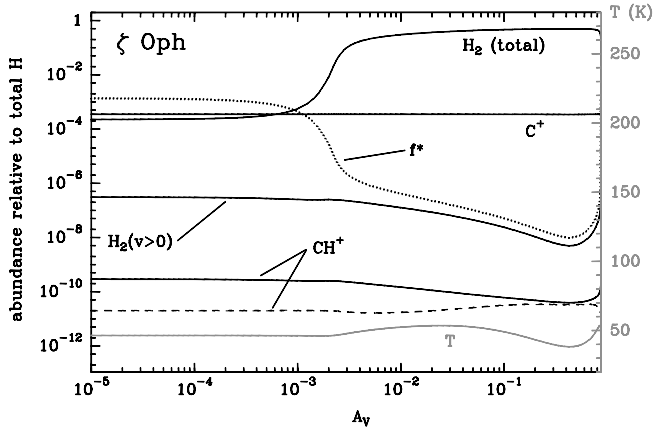}
\caption{Abundances relative to the total number of H nuclei in
$\zeta$ Oph for H$_2$, H$_2$ ($v>0$), C$^+$, and CH$^+$ as a
function of $A_V$ as given by the \emph{H$_2^*$ model} (dashed
line corresponds to the CH$^+$ abundance under the \emph{standard
model}). Also shown is the fraction of vibrationally excited H$_2$
$f^*$ (referred to the left axis) and the gas kinetic temperature
T (referred to the right axis).} \label{fig-zoph}
\end{figure}

Vibrationally excited molecular hydrogen is commonly observed in
PDRs, where the H$_2$ $v>0$ levels are excited by FUV
fluorescence, and in shocked gas where the excitation is mainly
collisional. Among these two types of regions the chemistry of
vibrationally excited H$_2$ is expected to be less important in
shocked regions since kinetic temperatures are usually very high
and the H$_2$ level populations are not as far from thermalization
as they are in PDRs \citep{bur92}. Hereafter we thus focus on the
case of PDRs to investigate the effects of the reactions of
vibrationally excited H$_2$ onto the gas chemical composition.

For that purpose we have utilized an updated version of the Meudon
PDR code\footnote{See \texttt{http://pdr.obspm.fr/PDRcode.html}},
a photochemical model of a plane parallel and stationary PDR
\citep{lep06,goi07,gon08}, to compute the chemical and physical
structure as a function of the cloud depth for various PDR-like
regions. Penetration of FUV radiation strongly depends on dust
grain properties through dust absorption and scattering of FUV
photons which, in addition, determine the efficiency of the
dominant heating mechanisms of the gas (photoelectric heating). In
this work we use standard dust properties, i.e. grain sizes follow
a power-law distribution \citep{mat77,goi07} and explicitly
determine the gas temperature gradient by solving the cloud
thermal balance (see \citealt{lep06}). We have used the chemical
network described in the previous section and the reactions of
excited H$_2$ shown in Table~\ref{table-h2-reac}. For the reaction
between H$_2$ and C$^+$ we adopt the rotational state-specific
rate constant expressions (No 2 in Table~\ref{table-h2-reac}) for
H$_2$ ($j\leq$7) and assume the Langevin value (No 3 in
Table~\ref{table-h2-reac}) for H$_2$ ($j>7$ or $v>0$), states for
which the reaction becomes exothermic. Under thermal conditions
this approach results in a value equivalent to the thermal rate
constant (No 1 in Table~\ref{table-h2-reac}), at worst a factor of
$\sim$2 larger in the high temperature range ($>$1000 K), where
the rate constant starts to be dominated by H$_2$ $v>0$ states.

We consider (de-)excitation of H$_2$ levels by collisions with
H$_2$, H, and He \citep{leb99}. Collisions with H$^+$ are also
included based on the recent work of \citet{hue08}, although they
are found to be not important. Concerning collisions with
electrons, the rates of vibrational de-excitation of H$_2$ $v=1
\rightarrow 0$ are typically a few 10$^{-10}$ cm$^3$ s$^{-1}$
(from the cross sections recommended by \citealt{yoo08}), which
implies a critical density of electrons of $\sim$ 10$^3$
cm$^{-3}$. In low density regions, such as diffuse clouds, the
electron density is much lower and electron-impact vibrational
excitation is negligible, although in denser regions with a high
ionization degree, such as planetary nebulae, electrons could play
a role in producing non-thermal H$_2$ vibrational populations. The
implementation of (de-)excitation of H$_2$ by collisions with
electrons is however not straightforward as only a few
state-to-state cross sections are available \citep{yoo08}.

We are mainly interested in investigating how important is the
effect of the chemistry of vibrationally excited H$_2$ in the
various regions. We thus have systematically run for the different
sources a model in which we adopt the thermal rate constant
expressions for all reactions of H$_2$ (\emph{standard model}) and
another one including the state-specific rate constants for the
reactions of excited H$_2$ detailed in Table~\ref{table-h2-reac}
(\emph{H$_2^*$ model}).

\begin{figure}
\includegraphics[angle=-90,scale=.46]{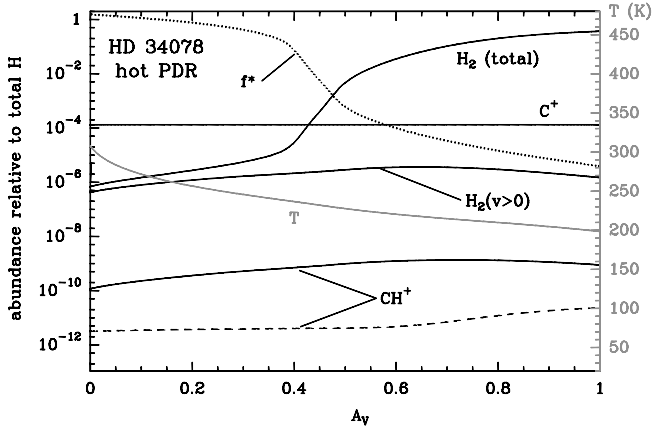}
\caption{Same as in Fig.~\ref{fig-zoph} but for the "hot PDR" of
HD 34078.} \label{fig-hd34078}
\end{figure}

\begin{deluxetable*}{l@{\hspace{0.2cm}}|r@{ }c@{ }l|r@{ }c@{ }l|r@{ }c@{ }l|r@{ }c@{ }l|r@{ }c@{ }l}
\tabletypesize{\scriptsize} \tablecaption{Calculated column
densities of selected species in the studied astronomical
sources\label{table-column-densities}} \tablewidth{0pc} \startdata
\hline
\multicolumn{1}{c}{}   & \multicolumn{3}{c}{$\zeta$ Oph}                     & \multicolumn{3}{c}{HD 34078 "hot PDR"}                      & \multicolumn{3}{c}{Orion Bar}                               & \multicolumn{3}{c}{NGC 7027}                                & \multicolumn{3}{c}{HD 176386} \\
\hline
$n_{\rm H}$(cm$^{-3}$) & \multicolumn{3}{c}{600}                             & \multicolumn{3}{c}{10$^4$}                                  & \multicolumn{3}{c}{10$^5$}                                  & \multicolumn{3}{c}{8$\times$10$^6$}                         & \multicolumn{3}{c}{3$\times$10$^3$} \\
$\chi$\tablenotemark{a}& \multicolumn{3}{c}{4$~||~$4}                        & \multicolumn{3}{c}{10$^4$$~||~$0}                           & \multicolumn{3}{c}{3$\times$10$^4$$~||~$0}                  & \multicolumn{3}{c}{5$\times$10$^3$\tablenotemark{b}$~||~$0} & \multicolumn{3}{c}{125\tablenotemark{c}$~||~$0} \\
$A_V^{\rm total}$      & \multicolumn{3}{c}{0.87}                            & \multicolumn{3}{c}{1.0}                                     & \multicolumn{3}{c}{9.0}                                     & \multicolumn{3}{c}{2.5}                                     & \multicolumn{3}{c}{0.9} \\
$T$(K)                 & \multicolumn{3}{c}{40$-$50}                         & \multicolumn{3}{c}{200$-$300}                               & \multicolumn{3}{c}{20$-$700}                                & \multicolumn{3}{c}{200$-$800}                               & \multicolumn{3}{c}{95$-$120} \\
\hline
\multicolumn{12}{c}{} \\
                       & \emph{standard} & $\rightarrow$ & \emph{H$_2^*$}    & \emph{standard} & $\rightarrow$ & \emph{H$_2^*$}            & \emph{standard} & $\rightarrow$ & \emph{H$_2^*$}            & \emph{standard} & $\rightarrow$ & \emph{H$_2^*$}            & \emph{standard} & $\rightarrow$ & \emph{H$_2^*$} \\
\hline
H              & 1.2$\times$10$^{20}$ & $\rightarrow$ & $\sim$               & 1.6$\times$10$^{21}$ & $\rightarrow$ & $\sim$               & 1.4$\times$10$^{21}$ & $\rightarrow$ & $\sim$               & 2.1$\times$10$^{21}$ & $\rightarrow$ & $\sim$               & 1.1$\times$10$^{20}$ & $\rightarrow$ & $\sim$               \\
H$_2$          & 7.6$\times$10$^{20}$ & $\rightarrow$ & $\sim$               & 1.6$\times$10$^{20}$ & $\rightarrow$ & $\sim$               & 7.7$\times$10$^{21}$ & $\rightarrow$ & $\sim$               & 1.3$\times$10$^{21}$ & $\rightarrow$ & $\sim$               & 8.8$\times$10$^{20}$ & $\rightarrow$ & 7.9$\times$10$^{20}$ \\
H$_2$($v>0$)   & 3.7$\times$10$^{13}$ & $\rightarrow$ & $\sim$               & 4.2$\times$10$^{15}$ & $\rightarrow$ & $\sim$               & 2.2$\times$10$^{16}$ & $\rightarrow$ & $\sim$               & 4.9$\times$10$^{17}$ & $\rightarrow$ & 4.2$\times$10$^{17}$ & 1.8$\times$10$^{14}$ & $\rightarrow$ & $\sim$               \\
C$^+$          & 5.6$\times$10$^{17}$ & $\rightarrow$ & $\sim$               & 2.5$\times$10$^{17}$ & $\rightarrow$ & $\sim$               & 7.5$\times$10$^{17}$ & $\rightarrow$ & $\sim$               & 4.6$\times$10$^{17}$ & $\rightarrow$ & 3.4$\times$10$^{17}$ & 3.6$\times$10$^{17}$ & $\rightarrow$ & $\sim$               \\
CH$^+$         & 5.2$\times$10$^{10}$ & $\rightarrow$ & 9.3$\times$10$^{10}$ & 1.4$\times$10$^{10}$ & $\rightarrow$ & 1.6$\times$10$^{12}$ & 4.1$\times$10$^{11}$ & $\rightarrow$ & 1.1$\times$10$^{13}$ & 5.7$\times$10$^{13}$ & $\rightarrow$ & 9.6$\times$10$^{13}$ & 5.1$\times$10$^{10}$ & $\rightarrow$ & 1.9$\times$10$^{11}$ \\
CH$_2^+$       & 1.3$\times$10$^{11}$ & $\rightarrow$ & $\sim$               & 3.3$\times$10$^{10}$ & $\rightarrow$ & 6.5$\times$10$^{11}$ & 4.6$\times$10$^{11}$ & $\rightarrow$ & 6.6$\times$10$^{12}$ & 1.8$\times$10$^{13}$ & $\rightarrow$ & 3.5$\times$10$^{13}$ & 1.3$\times$10$^{11}$ & $\rightarrow$ & 1.9$\times$10$^{11}$ \\
CH$_3^+$       & 2.3$\times$10$^{11}$ & $\rightarrow$ & $\sim$               & 3.8$\times$10$^{10}$ & $\rightarrow$ & 6.6$\times$10$^{11}$ & 1.9$\times$10$^{12}$ & $\rightarrow$ & 2.1$\times$10$^{13}$ & 5.0$\times$10$^{13}$ & $\rightarrow$ & 2.0$\times$10$^{14}$ & 4.7$\times$10$^{11}$ & $\rightarrow$ & 2.0$\times$10$^{11}$ \\
CH             & 1.7$\times$10$^{13}$ & $\rightarrow$ & $\sim$               & 2.6$\times$10$^{10}$ & $\rightarrow$ & 4.4$\times$10$^{11}$ & 4.3$\times$10$^{13}$ & $\rightarrow$ & 8.6$\times$10$^{13}$ & 7.7$\times$10$^{13}$ & $\rightarrow$ & 8.4$\times$10$^{13}$ & 3.2$\times$10$^{12}$ & $\rightarrow$ & 4.8$\times$10$^{12}$ \\
OH             & 1.1$\times$10$^{12}$ & $\rightarrow$ & $\sim$               & 3.7$\times$10$^{10}$ & $\rightarrow$ & 3.1$\times$10$^{11}$ & 4.6$\times$10$^{13}$ & $\rightarrow$ & $\sim$               & 2.5$\times$10$^{15}$ & $\rightarrow$ & $\sim$               & 1.1$\times$10$^{11}$ & $\rightarrow$ & 4.4$\times$10$^{11}$ \\
H$_2$O         & 1.0$\times$10$^{11}$ & $\rightarrow$ & $\sim$               & 2.5$\times$10$^{9}$  & $\rightarrow$ & 2.3$\times$10$^{10}$ & 4.4$\times$10$^{15}$ & $\rightarrow$ & $\sim$               & 1.0$\times$10$^{15}$ & $\rightarrow$ & $\sim$               & 1.1$\times$10$^{10}$ & $\rightarrow$ & 4.7$\times$10$^{10}$ \\
CO             & 8.0$\times$10$^{12}$ & $\rightarrow$ & $\sim$               & 8.2$\times$10$^{9}$  & $\rightarrow$ & 1.6$\times$10$^{11}$ & 1.4$\times$10$^{18}$ & $\rightarrow$ & $\sim$               & 2.2$\times$10$^{18}$ & $\rightarrow$ & $\sim$               & 2.0$\times$10$^{12}$ & $\rightarrow$ & 3.0$\times$10$^{12}$ \\
CN             & 4.4$\times$10$^{11}$ & $\rightarrow$ & $\sim$               & 1.0$\times$10$^{8}$  & $\rightarrow$ & 2.4$\times$10$^{9}$  & 2.1$\times$10$^{13}$ & $\rightarrow$ & 2.6$\times$10$^{13}$ & 3.2$\times$10$^{15}$ & $\rightarrow$ & $\sim$               & 1.7$\times$10$^{10}$ & $\rightarrow$ & 4.4$\times$10$^{10}$ \\
HCN            & 5.4$\times$10$^{8}$  & $\rightarrow$ & $\sim$               & 1.1$\times$10$^{6}$  & $\rightarrow$ & 2.4$\times$10$^{7}$  & 1.5$\times$10$^{13}$ & $\rightarrow$ & $\sim$               & 7.6$\times$10$^{15}$ & $\rightarrow$ & $\sim$               & 9.2$\times$10$^{7}$  & $\rightarrow$ & 1.7$\times$10$^{8}$  \\
\enddata
\tablecomments{The adopted elemental abundances for C, N, O, and S
are respectively 3.51$\times$10$^{-4}$, 7.5$\times$10$^{-5}$,
6.23$\times$10$^{-4}$, and 9.9$\times$10$^{-6}$ in $\zeta$ Oph;
1.32$\times$10$^{-4}$, 7.5$\times$10$^{-5}$,
3.19$\times$10$^{-4}$, and 1.85$\times$10$^{-5}$ in the "hot PDR"
of HD 34078; 1.4$\times$10$^{-4}$, 1.0$\times$10$^{-4}$,
3.0$\times$10$^{-4}$, and 2.8$\times$10$^{-5}$ in Orion Bar;
1.26$\times$10$^{-3}$, 1.91$\times$10$^{-4}$,
5.5$\times$10$^{-4}$, and 7.94$\times$10$^{-6}$ in NGC 7027; and
2.02$\times$10$^{-4}$, 5.07$\times$10$^{-5}$,
3.67$\times$10$^{-4}$, and 9.89$\times$10$^{-6}$ in HD 176386.}
\tablenotetext{a}{The notation $x~||~y$ refers to the FUV field
$x$ and $y$ at each side of the cloud.} \tablenotetext{b}{FUV
field at 0.0169 pc of a star with $T_*$ = 198000~K and $R_*$ =
0.075 R$_{\odot}$ radiating as a blackbody.} \tablenotetext{c}{FUV
field at 0.04 pc of a star with $T_*$ = 11300~K and $R_*$ = 2.55
R$_{\odot}$ radiating as a blackbody.}
\end{deluxetable*}

\subsection{Diffuse clouds: $\zeta$ Oph and HD 34078}

We first focus on a well known diffuse cloud such as $\zeta$ Oph,
whose parameters (see Table~\ref{table-column-densities}) have
been taken from \citet{van86}. In Fig.~\ref{fig-zoph} we show as a
function of $A_V$ the abundance of some species while the
calculated column densities are given in
Table~\ref{table-column-densities}. The largest effects of the
chemistry of vibrationally excited H$_2$ are seen at the edges of
the cloud ($A_V < 0.1$) where there is an important fraction of
H$_2$ ($v>0$), which gives rise to an increase in the abundance of
CH$^+$. Note that the sharp increase of $f^*$ (the ratio of
vibrationally excited to ground state H$_2$) up to $\sim$
10$^{-3}$ at the very outer edges ($A_V < 0.001$) is produced by
the decrease in the abundance of H$_2$ ($v =0$), relative to total
hydrogen, while that of H$_2$ ($v>0$) keeps roughly constant. In
this regime large $f^* \Delta k (T)$ ratios are attained (see
discussion at the end of Sec.~\ref{subsec-h2chem-basics}) and
reactions of vibrationally excited H$_2$ increase their importance
with respect to those of ground state H$_2$, but not necessarily
with respect to other processes (e.g. photodissociations). That is
why the jump of $f^*$ at the outer edge of $\zeta$ Oph is not
followed by the abundance of CH$^+$ (see Fig.~\ref{fig-zoph}).

In any case, as the fraction of H$_2$ ($v>0$) remains relatively
low ($f^*$ $<$ 10$^{-7}$) throughout most of the cloud, the total
column density of CH$^+$ is not dramatically affected. It
increases by about a factor of 2 (see Table
~\ref{table-column-densities}) but remains well below the observed
value of 2.9$\times$10$^{13}$ cm$^{-2}$ (see \citealt{van86}).
This relatively low fraction of vibrationally excited H$_2$ is in
agreement with astronomical observations of $\zeta$ Oph
\citep{fed95}, and implies that for quiescent diffuse clouds the
chemistry of vibrationally excited H$_2$ does not have dramatic
effects on the global chemical composition of the cloud.

Large fractions of vibrationally excited H$_2$ are not usually
observed in diffuse clouds, except for a few lines of sight: HD
37903 \citep{mey01} and HD 34078 \citep{boi05,boi09}, although its
presence seems to be related to a dense component exposed to a
high FUV field arising from nearby hot stars. In
Fig.~\ref{fig-hd34078} we show the results obtained for the dense
component of HD 34078, called the "hot PDR" by \citet{boi05},
whose parameters have been taken from these authors (see
Table~\ref{table-column-densities}). In this case, due to the
large FUV field, a high fraction of H$_2$ ($v>0$) is maintained
throughout the cloud. Thus, the reaction between H$_2$ ($v>0$) and
C$^+$ readily occurs affecting the abundance of various species
(see column densities in Table~\ref{table-column-densities}), most
notably CH$^+$ whose abundance is increased by two orders of
magnitude with respect to that obtained in the \emph{standard
model}, yet remaining below the observed value
(6.5$\times$10$^{13}$ cm$^{-2}$; see \citealt{boi05}). We note
that the differences between the column densities calculated in
our \emph{standard model} and those obtained by \citet{boi05} are
most likely due to differences in the chemical network utilized.
For instance, the larger column density of CH$^+$ obtained by them
is to a large extent due to the larger rate constant they adopt
for the radiative association between C$^+$ and H$_2$:
1.7$\times$10$^{-15}$ cm$^{3}$ s$^{-1}$ (upper limit derived for
para H$_2$ by \citealt{ger94}) versus the more conservative value
of 4$\times$10$^{-16}$($T/300$)$^{-0.2}$ cm$^{3}$ s$^{-1}$ adopted
by us (based on a calculation by \citealt{her85}).

We thus find that the reaction of H$_2$ ($v>0$) and C$^+$ is not
able by itself to explain the large CH$^+$ column densities
observed in diffuse and translucent clouds ($\geq$ 10$^{13}$
cm$^{-2}$). Nowadays, the most popular theories on the CH$^+$
formation in diffuse clouds invoke the existence of temporary
events, such as shocks \citep{pin86} or turbulence dissipation
\citep{god09}, which increase locally the gas kinetic temperature
and allow reaction (\ref{reac-h2+cplus}) to proceed. Another
possibility is that given by \citet{fed96}, who suggest that the
ion-neutral drift produced by Alfv\'en waves could result in a non
thermal rate enhancement of ion-neutral reactions, allowing CH$^+$
to be effectively formed through reaction (\ref{reac-h2+cplus}).
An enhanced FUV field has also been proposed as the ultimate cause
of CH$^+$ formation \citep{sno93}, hypothesis that is supported by
the fact that in diffuse clouds the column density of CH$^+$ is
well correlated with rotationally excited H$_2$ ($j \approx 3-5$)
\citep{lam86}, levels which are to a large extent pumped by FUV
fluorescence in diffuse clouds. An intense FUV field will indeed
bring H$_2$ molecules to vibrational excited states, which in turn
will react fast with C$^+$ to form CH$^+$, but when the various
magnitudes are evaluated through a PDR model under a plausible
physical scenario for a diffuse cloud it is found that the
abundance of H$_2$ ($v>0$) is not high enough to form CH$^+$ with
an abundance close to the the typical observed values, as shown in
the case of $\zeta$ Oph.

\subsection{Dense interstellar PDRs: Orion Bar}

\begin{figure}
\includegraphics[angle=-90,scale=.46]{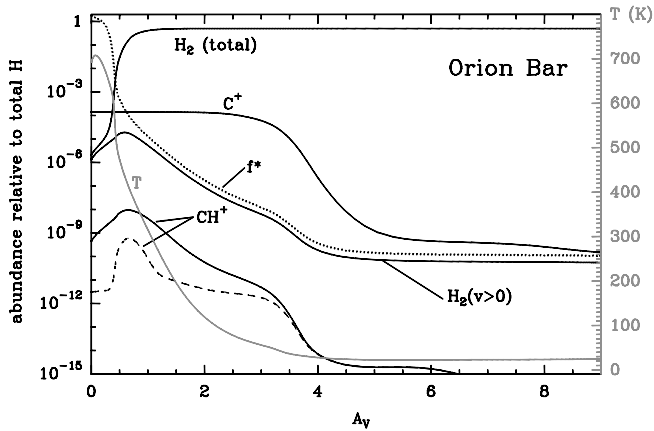}
\caption{Same as in Fig.~\ref{fig-zoph} but for the Orion Bar.}
\label{fig-orionbar}
\end{figure}

Dense and highly FUV-illuminated clouds are regions where the
chemistry of vibrationally excited H$_2$ is expected to play an
important role as the fraction of H$_2$ ($v>0$) is usually large.
Here we consider the case of one of the most archetypical dense
PDRs which is the Orion Bar, whose parameters (see
Table~\ref{table-column-densities}) have been taken from
\citet{van09}. In this case the combination of a high FUV
radiation field and a high density ($n_{\rm H}/\chi$ $\sim$ 3)
results in a fraction of H$_2$ ($v>0$) in excess of 10$^{-6}$ in
the region $A_V = 0.5-1.5$, where hydrogen is still mostly
molecular and the gas kinetic temperature is just of a few
hundreds of degrees Kelvin (see Fig.~\ref{fig-orionbar}). Thus,
the reaction H$_2$ ($v>0$) + C$^+$ has indeed a great impact on
the abundance of CH$^+$ which increases by almost 2 orders of
magnitude with respect to that obtained in the \emph{standard
model}.

The existence of an important amount of vibrationally excited
H$_2$ in the Orion Bar is supported by astronomical observations
\citep{van96}. On the other hand, CH$^+$ has not yet been observed
but its pure rotational lines lie in the wavelength range of the
Herschel Space Observatory. If, as indicated by our model, the
formation of CH$^+$ occurs by the reaction of H$_2$ ($v>0$) and
C$^+$, then the spatial distribution of CH$^+$ emission should be
well correlated with that of the $v$ = 1$\rightarrow$0 S(1) line
of H$_2$ observed by \citet{van96}

\subsection{Planetary nebulae: NGC 7027}

\begin{figure}
\includegraphics[angle=-90,scale=.46]{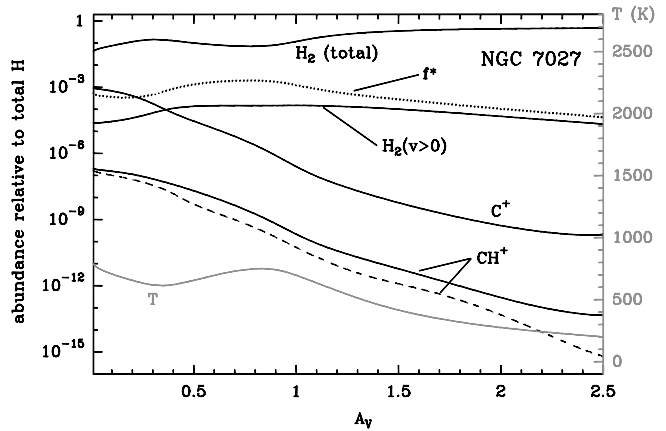}
\caption{Same as in Fig.~\ref{fig-zoph} but for NGC 7027.}
\label{fig-ngc7027}
\end{figure}

Planetary nebulae are also sources where the chemistry of
vibrationally excited H$_2$ can be important since the
circumstellar gas is normally exposed to a very high FUV flux
which emanates from the hot central white dwarf. A prototype of
this kind of objects is NGC 7027, which harbors one of the hottest
known stars ($T_*$ $\sim$ 200,000~K). To model this source we have
adopted the physical scenario utilized by \citet{has00} focusing
on the dense region (see Table~\ref{table-column-densities}). As
shown in Fig.~\ref{fig-ngc7027} a large fraction of H$_2$ ($v>0$)
is present in the circumstellar gas of this source ($f^*$ $\sim$
10$^{-4}$$-$10$^{-3}$), and CH$^+$ is readily formed by the
reaction of H$_2$ ($v>0$) and C$^+$ in the regions located close
to the star, where C$^+$ reaches its maximum abundance. In these
regions the reaction H$_2$ ($v=0$) + C$^+$ is also an important
source of CH$^+$ due to the high temperatures attained.

From the observational point of view the existence of
vibrationally excited H$_2$ in NGC 7027 is well established from
ISO observations \citep{ber01}, as is the presence of CH$^+$ with
a column density of 0.8$-$2.5$\times$10$^{14}$ \citep{cer97}, in
agreement with the value we get.

\subsection{Circumstellar disks: HD 176386}

\begin{figure}
\includegraphics[angle=-90,scale=.46]{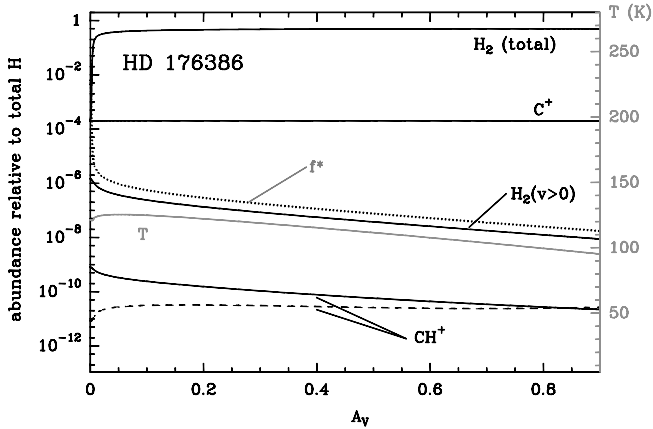}
\caption{Same as in Fig.~\ref{fig-zoph} but for HD 176386.}
\label{fig-hd176386}
\end{figure}

Finally, we treat also the case of circumstellar disks surrounding
young stars of T Tauri and Herbig Ae/Be type, where the outer disk
surface located at just a few AU of the central star is exposed to
the strong stellar UV flux. Hot molecular hydrogen both
rotationally and vibrationally excited, has been observed in
various sources of this type \citep{mar08}. To investigate the
importance of the chemistry of vibrationally excited H$_2$ in such
environments we take the parameters derived for the warm H$_2$
emitting gas observed by \citet{mar08} toward the B9 star HD
176836 (see Table~\ref{table-column-densities}). As shown in
Fig.~\ref{fig-hd176386} the fraction of vibrationally excited
H$_2$ stays around 10$^{-7}$$-$10$^{-6}$ and so the abundance of
CH$^+$ is increased by almost one order of magnitude when
reactions with excited H$_2$ are included, although in the case of
HD 176386 the total column density (see
Table~\ref{table-column-densities}) remains probably too low for
being detected. Some other species experience moderate abundance
enhancements as a consequence of the activation of the reaction
between H$_2$ ($v>0$) and C$^+$.

\section{Summary}

In this paper we have investigated the impact of various chemical
reactions involving vibrationally excited H$_2$ which have been
studied either experimentally or theoretically (concretely those
with C$^+$, He$^+$, O, OH, and CN) on the chemical composition of
various astronomical regions. Among the reactions considered here
that of H$_2$ ($v>0$) with C$^+$ stands out as the most important
one. This reaction becomes the main formation route of the
reactive cation CH$^+$ and controls the abundance of some other
related species in moderately warm astronomical environments with
fractions of vibrationally excited H$_2$ in excess of some
10$^{-6}$, conditions which are found to occur in some types of
PDR-like regions.

The importance of the chemistry of vibrationally excited H$_2$ has
been investigated through PDR models in the diffuse clouds $\zeta$
Oph and HD 34078, the Orion Bar, the carbon-rich protoplanetary
nebula NGC 7027, and the circumstellar disk HD 176386. The study
of $\zeta$ Oph indicates that the fraction of H$_2$ ($v>0$)
present ($<$ 10$^{-7}$) is insufficient to form CH$^+$ with the
abundance typically observed in diffuse and translucent clouds. On
the other hand, in dense and highly FUV illuminated clouds, such
as the hot PDR of HD 34078, the Orion Bar and NGC 7027, reactions
of vibrationally excited become crucial to determine the global
chemical composition. In particular, the reaction H$_2$ ($v>0$) +
C$^+$ becomes the dominant synthetic pathway to CH$^+$ and make
the related species CH$_2^+$ and CH$_3^+$ to reach similar or
slightly lower abundances. These two cations however have not yet
been observed in any astronomical region $-$ the spectroscopic
data of CH$_2^+$ is poorly known (see e.g. \citealt{pol07}) while
CH$_3^+$ lacks a permanent dipole moment.

Other reactions of vibrationally excited H$_2$ not yet studied
either experimentally or theoretically, and thus not included
here, could become important in regulating the chemical balance of
interstellar clouds. For instance, the reactions H$_2$ + C
$\rightarrow$ CH + H, H$_2$ + S $\rightarrow$ SH + H, and H$_2$ +
S$^+$ $\rightarrow$ SH$^+$ + H are endothermic by 0.99 eV ($\sim$
11500~K), 0.83 eV ($\sim$ 9600 K), and 0.86 eV ($\sim$ 9860 K)
respectively, and their rate constants have an exponential term of
about the value of the endothermicity. This may indicate that,
analogously to reaction (\ref{reac-h2+cplus}), the internal energy
of excited H$_2$ would be effectively used to diminish or overcome
the activation barrier so that for H$_2$ ($v>1$), all these
reactions being exothermic, they would proceed at about the
collision limit. Experimental or theoretical information on these
or other potentially important reactions of vibrationally excited
H$_2$ would be of great interest.

In any case we encourage astrochemists to include the reactions of
excited H$_2$ for which kinetic data is available (see
Table~\ref{table-h2-reac}) when modelling the chemistry of regions
with an important fraction of vibrationally excited H$_2$. At
least the reaction between H$_2$ ($v>0$) and C$^+$ clearly has an
important impact on the global chemical composition of such
regions.

\acknowledgments

We thank the referee for constructive comments on this article.
M.A. is supported by a \textit{Marie Curie Intra-European
Individual Fellowship} within the European Community 7th Framework
Programme under grant agreement n$^{\circ}$ 235753. J.R.G. was
supported by a Ram\'on y Cajal research contract from the Spanish
MICINN and co-financed by the European Social Fund.


\begin{thebibliography}{}

\bibitem[Ag\'undez \& Cernicharo(2006)]{agu06} Ag\'undez, M. \& Cernicharo, J. 2006, \apj, 650, 374
\bibitem[Asplund et al.(2009)]{asp09} Asplund, M., Grevesse, N., Sauval, A. J., \& Scott, P. 2009, \araa, 47, 481
\bibitem[Bernard Salas et al.(2001)]{ber01} Bernard Salas, J., Pottasch, S. R., Beintema, D. A., \& Wesselius, P. R. 2001, \aap, 367, 949
\bibitem[Boiss\'e et al.(2005)]{boi05} Boiss\'e, P., Le Petit, F., Rollinde, E., et al. 2005, \aap, 429, 509
\bibitem[Boiss\'e et al.(2009)]{boi09} Boiss\'e, P., Rollinde, E., Hily-Blant, P., et al. 2009, \aap, 501, 221
\bibitem[Burton(1992)]{bur92} Burton, M. G. 1992, \ajp, 45, 463
\bibitem[Cernicharo et al.(1997)]{cer97} Cernicharo, J., Liu, X.-W., Gonz\'alez-Alfonso, E., et al. 1997, \apjl, 483, L65
\bibitem[Chase(1998)]{cha98} Chase, M. W. 1998, NIST-JANAF Thermochemical Tables, 4th ed., \jpcrd, Monograph n 9
\bibitem[Dabrowski(1984)]{dab84} Dabrowski, I. 1984, \cjp, 62, 1639
\bibitem[Draine(1978)]{dra78} Draine, B. T. 1978, \apjs, 36, 595
\bibitem[Federman et al.(1995)]{fed95} Federman, S. R., Cardelli, J. A., van Dishoeck, E. F., Lambert, D. L., \& Black, J. H. 1995, \apj, 445, 325
\bibitem[Federman et al.(1996)]{fed96} Federman, S. R., Rawlings, J. M. C., Taylor, S. D., \& Williams, D. A. 1996, \mnras, 279, L41
\bibitem[Freeman \& Williams(1982)]{fre82} Freeman, A. \& Williams, D. A. 1982, \apss, 83, 417
\bibitem[Garrod et al.(2003)]{gar03} Garrod, R. T., Rawlings, J. M. C., \& Williams, D. A. 2003, \apss, 286, 487
\bibitem[Gerlich et al.(1987)]{ger87} Gerlich, D., Disch, R., \& Scherbarth, S. 1987, \jcp, 87, 350
\bibitem[Gerlich \& Rox(1989)]{ger89} Gerlich, D. \& Rox, T. 1989, \zfpd, 13, 259
\bibitem[Gerlich(1994)]{ger94} Gerlich, D. 1994, in AIP Conf. Proc. 312, Molecules and Grains in Space, ed. I. Nenner (New York: AIP), 489
\bibitem[Godard et al.(2009)]{god09} Godard, B., Falgarone, E., \& Pineau des F$\hat{\rm o}$rets, G. 2009, \aap, 495, 847
\bibitem[Goicoechea \& Le Bourlot(2007)]{goi07} Goicoechea, J. R. \& Le Bourlot, J. 2007, \aap, 467, 1
\bibitem[Gonz\'alez Garc\'ia et al.(2008)]{gon08} Gonz\'alez Garc\'ia, M., Le Bourlot, J., Le Petit, F., \& Roueff, E. 2008, \aap, 485, 127
\bibitem[Hasegawa et al.(2000)]{has00} Hasegawa, T., Volk, K., \& Kwok, S. 2000, \apj, 532, 994
\bibitem[Herbst et al.(1985)]{her85} Herbst, E. 1985, \apj, 291, 226
\bibitem[Hierl et al.(1997)]{hie97} Hierl, P. M., Morris, R. A., \& Viggiano, A. A. 1997, \jcp, 106, 10145
\bibitem[Hollenbach \& McKee(1989)]{hol89} Hollenbach, D. \& McKee, C. F. 1989, \apj, 342, 306
\bibitem[Huestis(2008)]{hue08} Huestis, D. L. 2008, \planss, 56, 1733
\bibitem[Jones et al.(1986)]{jon86} Jones, M. E., Barlow, S. E., Ellison, G. B., \& Ferguson, E. E. 1986, \cpl, 130, 218
\bibitem[Jura(1975)]{jur75} Jura, M. 1975, \apj, 197, 575
\bibitem[Lambert \& Danks(1986)]{lam86} Lambert, D. L. \& Danks, A. C. 1986, \apj, 303, 401
\bibitem[Le Bourlot(1991)]{leb91} Le Bourlot, J. 1991, \aap, 242, 235
\bibitem[Le Bourlot et al.(1999)]{leb99} Le Bourlot, J., Pineau des F$\hat{\rm o}$rets, G., \& Flower, D. R. 1999, \mnras, 305, 802
\bibitem[Le Petit et al.(2006)]{lep06} Le Petit, F., Nehm\'e, C., Le Bourlot, J., \& Roueff, E. 2006, \apjs, 164, 506
\bibitem[Lee et al.(1996)]{lee96} Lee, H.-H., Herbst, E., Pineau des F$\hat{\rm o}$rets, G., Roueff, E., \& Le Bourlot, J. 1996, \aap, 311, 690
\bibitem[Martin-Za\"{\i}di et al.(2008)]{mar08} Martin-Za\"{\i}di, C., Deleuil, M., Le Bourlot, J., et al. 2008, \aap, 484, 225
\bibitem[Mathis et al.(1977)]{mat77} Mathis, J. S., Rumpl, W., Nordsieck, K. H. 1977, \apj, 217, 425
\bibitem[Meyer et al.(2001)]{mey01} Meyer, D. M., Lauroesch, J. T., Sofia, U. J., Draine, B. T., \& Bertoldi, F. 2001, \apjl, 553, L59
\bibitem[Olkhov \& Smith(2007)]{olk07} Olkhov, R. V. \& Smith, I. W. M. 2007, \jcp, 126, 134314
\bibitem[Park et al.(2006)]{par06} Park, I. H., Wakelam, V., \& Herbst, E. 2006, \aap, 449, 631
\bibitem[Pineau des F$\hat{\rm o}$rets et al.(1986)]{pin86} Pineau des F$\hat{\rm o}$rets, G., Flower, D. R., Hartquist, T. W., \& Dalgarno, A. 1986, \mnras, 220, 801
\bibitem[Polehampton et al.(2007)]{pol07} Polehampton, E. T., Baluteau, J.-P., Swinyard, B. M., et al. 2007, \mnras, 377, 1122
\bibitem[Schauer et al.(1989)]{sch89} Schauer, M. M., Jefferts, S. R., Barlow, S. E., \& Dunn, G. H. 1989, \jcp, 91, 4593
\bibitem[Snow(1993)]{sno93} Snow, T. P. 1993, \apjl, 402, L73
\bibitem[Stecher \& Williams(1972)]{ste72} Stecher, T. P. \& Williams, D. A. 1972, \apjl, 177, L141
\bibitem[Sternberg \& Dalgarno(1995)]{ste95} Sternberg, A. \& Dalgarno, A. 1995, \apjs, 99, 565
\bibitem[Sultanov \& Balakrishnan(2005)]{sul05} Sultanov, R. A. \& Balakrishnan, N. 2005, \apj, 629, 305
\bibitem[Tielens \& Hollenbach(1985)]{tie85} Tielens, A. G. G. M. \& Hollenbach, D. 1985, \apj, 291, 722
\bibitem[Truong(1995)]{tru95} Truong, T. N. 1995, \jcp, 102, 5335
\bibitem[van der Werf et al.(1996)]{van96} van der Werf, P. P., Stutzki, J., Sternberg, A., \& Krabbe, A. 1996, \aap, 313, 633
\bibitem[van der Wiel et al.(2009)]{van09} van der Wiel, M. H. D., van der Tak, F. F. S., Ossenkopf, V., et al. 2009, \aap, 498, 161
\bibitem[van Dishoeck \& Black(1986)]{van86} van Dishoeck, E. F. \& Black, J. H. 1986, \apjs, 62, 109
\bibitem[Wagner \& Graff(1987)]{wag87} Wagner, A. F. \& Graff, M. M. 1987, \apj, 317, 423
\bibitem[Woodall et al.(2007)]{woo07} Woodall, J., Ag\'undez, M., Markwick-Kemper, A. J., \& Millar, T. J. 2007, \aap, 466, 1197
\bibitem[Yoon et al.(2008)]{yoo08} Yoon, J.-S., Song, M.-Y., Han, J.-M., et al. 2008, \jpcrd, 37, 913
\bibitem[Zellner \& Steinert(1981)]{zel81} Zellner, R. \& Steinert, W. 1981, \cpl, 81, 568
\bibitem[Zhu et al.(1998)]{zhu98} Zhu, W., Zhang, J. Z. H., Zhang, Y. C., et al. 1998, \jcp, 108, 3509

\end{thebibliography}
\end{document}